\documentclass{jfm}
\usepackage{graphicx}
\usepackage{epstopdf, epsfig}
\usepackage{amsmath}
\usepackage{bm}
\usepackage{ulem}
\usepackage{xfrac}
\usepackage{natbib}

\usepackage{color}
\newcommand{\ew}[1]{\textcolor{black}{#1}}

\shorttitle{Experimental structure functions}
\shortauthor{E-W. Saw,P. Debue,D. Kuzzay, F. Daviaud and B.  Dubrulle}

\title{On the universality of anomalous scaling exponents of structure functions in turbulent flows}
\author{E-W. Saw\aff{1,2}, P. Debue \aff{1}, D. Kuzzay\aff{1}, F. Daviaud\aff{1},  \and B.  Dubrulle\aff{1}$^{\corresp{\email{berengere.dubrulle@cea.fr}}}$}

\affiliation{\aff{1}{SPEC/IRAMIS/DSM, CEA, CNRS, University Paris-Saclay, CEA Saclay, 91191 Gif-sur-Yvette, France.
\aff{2}{School of Atmospheric Sciences, Sun Yat-Sen University, Guangzhou, China}}}

\begin{document}

\maketitle

\begin{abstract}
All previous experiments in open turbulent flows (e.g. downstream of grids, jet and atmospheric boundary layer) have produced quantitatively consistent values for the scaling exponents of velocity structure functions \citep{ Anselmet1984, Stolovitzky1993, Arneodo96}. The only measurement \ew{of scaling exponent at high order ($> 6$) }in closed turbulent flow (von K\'arm\'an swirling flow) using \ew{Taylor's frozen flow hypothesis}, however, produced scaling exponents that are significantly smaller, suggesting that the universality of these exponents are broken with respect to change of large scale geometry of the flow. Here, we report measurements of longitudinal structure functions of velocity in a von K\'arm\'an setup without the use of Taylor-hypothesis. The measurements are made using Stereo Particle Image Velocimetry at 4 different ranges of spatial scales, in order to observe a combined inertial subrange spanning roughly one and a half order of magnitude. We found scaling exponents (up to 9th order) that are consistent with values from open turbulent flows, suggesting that they might be in fact universal.  
  
 \end{abstract}


\section{Introduction}
In the classical Kolmogorov-Richardson  picture of turbulence, a  turbulent flow is characterized by a hierarchy of self-similar scales. This picture becomes increasingly 
inaccurate at smaller and smaller scales, where intermittent burst of energy dissipation and transfers take place  \citep{K62}. A classical quantification of such intermittency
is via the 
scaling properties of the structure functions, built as successive moments $ S_n(r)=<\vert\delta u_r\vert^n>$ of $\delta u_r={\bf r}\cdot({\bf u}(x+r)-{\bf u}(x))/r$ the longitudinal velocity increments over a distance $r$.
%\ew{the anomalous deviation of scaling properties of the velocity structure functions from the simplistic \cite{K41a} scaling}. The velocity structure functions is defined as successive moments $ S_n(r)=<\vert\delta u_L\vert^n>$ of $\delta u_L={\bf r}\cdot({\bf u}(x+r)-{\bf u}(x))/r$ the longitudinal velocity increments over a distance $r$. 
In numerical simulations, the longitudinal increments are easily accessible over the whole range of scale of the
simulations, but the scaling ranges and the maximal order $n$ are limited by numerical resources. In experiments, large Reynolds numbers and large statistics
are easily accessible, but the computation of $S_n(r)$ faces practical challenges. One point velocity measurements based e.g. on hot wire or LDV techniques- provide
time-resolved measurements over 3 or 4  decades, that can be used to compute the  structure functions only via the so-called Taylor's frozen flow hypothesis $r=U\Delta t $, where $U$ is the mean flow velocity at the probe location. This rules out the use of this method in ideal homogeneous isotropic turbulence, where $U=0$. Most of the experimental reports on structure function scalings relied on this method and not surprisingly most of these results are from turbulent flows with open geometry (e.g. turbulence dowmstream of grids, jets and atmospheric boundary layer) where there is a strong mean flow. The common practice is to keep the ratio $u'/U$ small, preferably less than $10\%$ ($u'$ being the standard deviation of the velocity). A summary of these results could be found in e.g.  \citep{Arneodo96}. For closed turbulent flows such as the von K\'arm\'an swirling flows, the best attempts \citep{Maurer94, Belin1996} were to place the point measurement probe at locations where $U$ is strongest, specifically where $U$ has a substantial azimuthal component. Even then, one had to resort to measurements with ratio of $u'/U$  up to $38\%$. At the same time, it is not understood how the curve geometry of the mean flow profile affects the validity of Taylor-hypothesis (this concern does not appear in open flows where $\vec{U}$ is predominantly rectilinear). Nevertheless, these experiments discerned power laws regime of the velocity structure functions of comparable quality with the results from open turbulence. When the various results on the scaling exponents were compared by \cite{Arneodo96}, the result from von K\'arm\'an flows gave values distinctly lower than those of open turbulence (which are among them selves consistent). This difference, among other reasons, had prompt suggestion that different classes of flows possess different sets of exponent  \citep{Sreenivasan97}. \ew{Besides, von K\'arm\'an flow, another branch of results on closed turbulent flow of the type Couette-Taylor was carried out, also utilizing Taylor's hypothesis,  by \cite{Lewis99} and \cite{Huisman2013}. Remarkably Lewis et al.'s results on the exponents beyond order-6 were also consistently lower than the open flow results and their highest Reynolds number measurements were remarkably close to that of \cite{Belin1996}. Huisman et al. reported results up to order-6 that suggested universality of the exponents with respect to changing large scale symmetry (by changing ratio of rotation of the cylinders) and to Reynolds number.}   Here using a technique that does not rely on Taylor-hypothesis, we report scaling exponents from von K\'arm\'an flows that are consistent with results from open flows  \citep{Arneodo96, Stolovitzky1993, Anselmet1984}, as well as results from numerical simulations  \citep{Gotoh02,Ishihara09} and the theory of \cite{She94}. We close this section by noting that \cite{Pinton94} had attempted to apply their original ``local Taylor-hypothesis" on a von K\'arm\'an experiment but only reported scaling exponent of up to sixth order which they concluded as being consistent with other results of the open flows and thus also with our results.   

\section{Overview of methods}
\begin{figure}
\includegraphics[width=.35\textwidth]{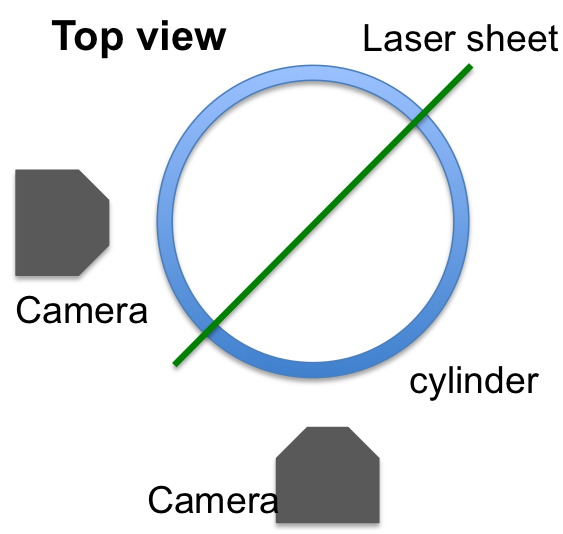}
\centering
\caption{\label{setup} Sketch of the apparatus. Velocity fields are observed in a two dimensional area on a laser sheet centered on the center of the experiment (equal distanced from the impellers and on the axis the cylindrical tank) . Two cameras view the observation area at angles roughly 45 degrees from the laser sheet.}
\vspace{-0pt}
\end{figure}

Direct measurement of spatial increments of velocity can be obtained via Particle Image Velocimetry method (PIV). A detailed description of the setup has been previously provided in \citep{Saw2016}, here we provide a concise description. The fluid is seeded with \ew{hollow glass particles (Dantec Dynamics) with mass density of $1.4gcm^{-3}$ and size $10-30\mu m$, giving particle Kolmogorov scale Stokes number, in order of increasing flow Reynolds numbers, of the order of $10^{-4}$ to $10^{-2}$, while the settling parameter i.e. ratio of Stokes to Froude number of the order of $10^{-3}$ or smaller}. \ew{The particles are illuminated by a thin laser sheet ($1mm$ thick) in center of the cylindrical tank (see Figure~\ref{setup} for a sketch). Two cameras, viewing at oblique angle from either side of the laser sheet,} take successive snapshots of the flow. The velocity field is then reconstructed \ew{across the quasi-two-dimensional laser sheet} using peak correlation performed over small interrogation windows. This methods provides measurements of \ew{3 velocity components on a two dimensional grid. However, limited width of the measured velocity field} (due to finite camera sensor size), coarse-graining of the reconstruction methods and optical noises usually limit the range of accessible scales, making the determination of the power law regime ambiguous, thus limiting the accuracy of measured scaling properties of the structure functions.
As we discuss in the present communication, these limitations can be overcame by combining multi-scale imaging and a universality hypothesis. In the original Kolmogorov self-similar theory (K41), for any $r$ in the inertial range, $ S_n(r)=C_n \epsilon^{n/3} r^{n/3}$, where $\epsilon$ is the (global) average energy dissipation and $C_n$ is a $n$-dependent constant. In such a case, the function $S_n(r)/ (\epsilon\eta)^{n/3}$ is a universal function (a power-law) of 
$r/\eta$, where $\eta=(\nu^3/\epsilon)^{1/4}$ is the Kolmogorov scale and $\nu$ is the liquid's kinematic viscosity. Such scaling is the only one compatible with the hypothesis that $r$ and $\epsilon$ are the only characteristic quantities in the inertial range. Following \cite{K62}, one can take into account possible breaking of the global self-similarity by assuming that there exists an additional characteristic  scale $\ell_0$ that matters in the inertial range, so that $S_n(r)=C_n \epsilon^{n/3} r^{n/3} F_n(r/\ell_0)$. If there exists a range of scale where $F_n(x)\sim x^{\alpha_n}$, then one can write $S_n(r)=C_n \epsilon^{n/3} \ell_0^{-(\zeta_n-n/3)}r^{\zeta_n}$, where $\zeta_n=n/3+\alpha_n$. In such a case, $G_n(r)=S_n(r) /(\epsilon\eta)^{n/3} (\ell_0/\eta)^{\zeta_n-n/3}$ is a universal function (a power-law) of $r/\eta$. 
 %This kind of universal shapes have been derived in the past to fit the structure functions by expressions that are universally valid from the dissipative scale to the inertial range  \citep[see][]{sirovich,stolovitzky}. They were found to give quite accurate descriptions of experimental and numerical structure functions. 
 Here, we use this to rescale our measurements taken in the same geometry,  but with different $\epsilon$ and $\eta$.  We may then collapse them in the inertial range into a single (universal) structure function by considering $G_n(r)$ as a function of $r/\eta$. The quality of the collapse depends crucially on the intermittency parameter $(\alpha_n =\zeta_n-n/3)$ for large value of $n$ , as $\alpha_n$ increases with $n$: a bad choice of $\alpha_n$ results in a strong mismatch of two measurements taken at same $r/\eta$ but  different $\eta$. Moreover, the global slope (in a log-log plot) of the collapsed data representing $G_n(r/\eta)$ can also be used to compute the effective scaling exponent $\zeta_n$ (and also $\alpha_n$), therefore providing a strong consistency check for the estimated value of $\zeta_n$. The bonus with the computation of the global slope is that by a proper choice of $\epsilon$ (which gives $\eta$ via $(\nu^3/\epsilon)^{1/4}$ ) and $\ell_0$, it can be performed over a wider inertial range, therefore allowing a more precise estimate of $\zeta_n$. In the sequel, we examine the effectiveness of this approach.
 
 %Overall, the collapse method from different PIV measurements, at different scale and different $\epsilon$ and $\eta$ therefore provides accurate estimates of the intermittency corrections, as we now show in a turbulent swirling flow.

\section{Experimental flow field}\label{sec:exp_flow}

We use an experimental von K\'arm\'an set-up that has been especially designed to allow for long time (up to hours) measurement of flow velocity to accumulate enough statistics for reliable data analysis. Turbulence is generated by two counter rotating impellers, in a cylindrical vessel of radius $R=10\,cm$ filled with water-glycerol mixtures  \cite[see][for a detailed description]{Saw2016} . We perform our measurements in the center region of the flow, with area of views of $4\times 4 \,cm^2$ \ew{(except in one case $20\times 20\,cm^2$, i.e. case D in Table \ref{table:param}), located on a meridian plane, around the symmetry point of the experimental set-up (see Fig.~\ref{setup} ).} At this location, a shear-layer induced by the differential rotation produces strong turbulent motions.
% that are  thought to be representative of classical homogeneous, isotropic turbulence   \citep{pinton,Tabeling}. 
Previous study of intermittency in such a set-up has been performed via one-point velocity measurements (hot-wire method) located above the shear layer or near the outer cylinder, where the mean velocity is non-zero \citep{Maurer94, Belin1996}. The scaling properties of structure functions up to $n=6$ were performed by measuring the scaling exponents $\zeta_n$ via $S_n(r)\sim r^{\zeta_n}$, using assumption of Taylor's frozen turbulence hypothesis and the extended self-similarity (ESS)  technique \citep{Benzi93}. These resulting $\zeta_n$ values were significantly lower than those from open turbulent flows \citep{Arneodo96}. The $\zeta_n$ values from open turbulence flows summarized in \cite{Arneodo96} are reproduced here (in the next section) for comparison.
Here, we use SPIV measurements of velocities to compute longitudinal structure function up to the ninth order without using Taylor-hypothesis \ew{(only velocity components in the measurement plane are used)}.  Our multi-scale imaging provides the possibility to access scales of the order (or smaller than) the dissipative scale, in a fully developed turbulent flow. The dissipative scale $\eta$ is proportional to the experiment size, and decreases with increasing Reynolds number. Tuning of the dissipative scale is achieved through viscosity variation, using different fluid mixtures of glycerol and water. Combined with variable optical magnifications, we may then adjust our resolution, to span a range of scale between $\eta$ to almost $10^4$ $\eta$, achieving roughly 1.5 decades of inertial range. 
Table \ref{table:param} summarizes the parameters corresponding to the different cases. All cases are characterized by the same value of non-dimensional global energy dissipation $\epsilon_g=0.045$ \ew{(non-dimensionalized using radius of tank $R$ and $2\pi F$, where F is the frequency of the impellers)}, measured through independent torque acquisitions. However, since the von K\'arm\'an flow is globally inhomogeneous, the local non-dimensional energy dissipation may vary from cases to cases \citep{Kuzzay15}, and has to be estimated using local measurements, as we detail below.

\begin{table}
  \begin{center}
\def~{\hphantom{0}}
 \begin{tabular}{lcccccccc}%
  Case&F (Hz)&Glycerol content &$Re$ &$\eta$ (mm)& $\Delta x$ (mm) &$\epsilon_g$ &$\epsilon_v$\\
%\hline 
%\hline
A & 1.2 &$59\%$ & $6\times 10^3$&$0.37$&$0.24$&$0.045$&$0.0275$\\
%\hline
B &1 &$0\%$ (at $5^o$C) & $6\times 10^4$&$0.0775 $&$0.48$&$0.045$&$0.0413$\\
%\hline
C &5 &$0\%$ & $3\times 10^5$&$0.0162$&$0.24$&$0.045$&$0.0502$\\
%\hline
D &5 &$0\%$ & $3\times 10^5$&$0.0193$&$2.4$&$0.045$&$0.0254$\\
%\hline
%E & 5 &$0\%$ & $3\times 10^5$&$0.02$&$3.4 $&$xx$&$xx$\\
%\hline \hline
\end{tabular}
\caption{Parameter space describing the  cases considered in this paper. $\Delta x$ is the spatial resolution of our measurements. $\epsilon_g$ is the global non-dimensional energy dissipation \ew{(non-dimensionalized using radius of tank ($R$) and $2\pi F$, where F is the frequency of the impellers)} measured through torques, while $\epsilon_v$ is the local non-dimensional energy dissipation estimated via the second order structure function (see the text on how they are estimated). $\eta$ is the Kolmogorov scale (estimated using $\nu$ and $\epsilon_v$). Liquid temperature is at $20^o$C unless otherwise specified.}
 \label{table:param}
\end{center}
\end{table}

\section{Results}\label{sec:results}
\subsection{Velocity increments and structure functions}
\begin{figure}
\includegraphics[width=.495\textwidth]{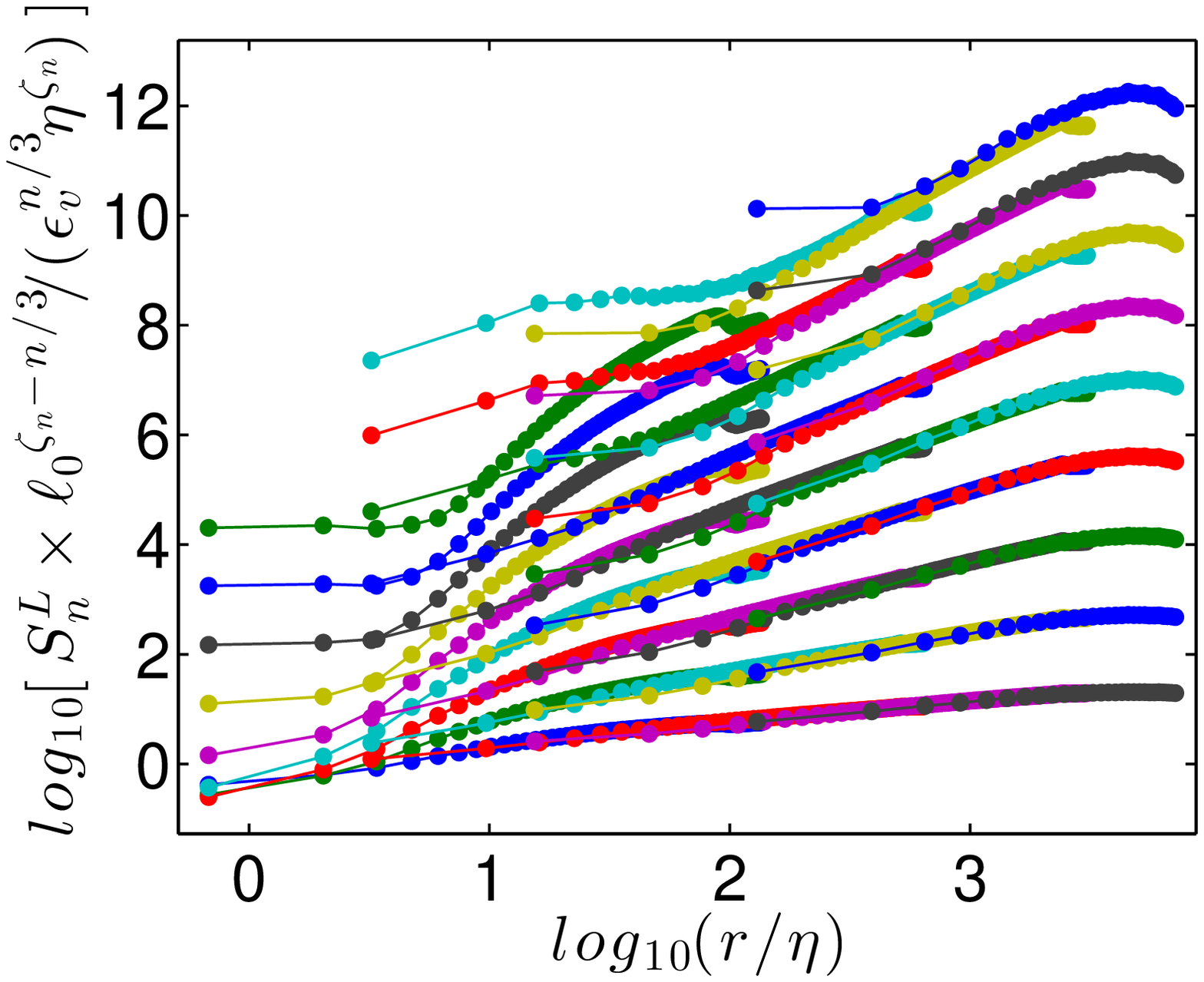}
\includegraphics[width=.495\textwidth]{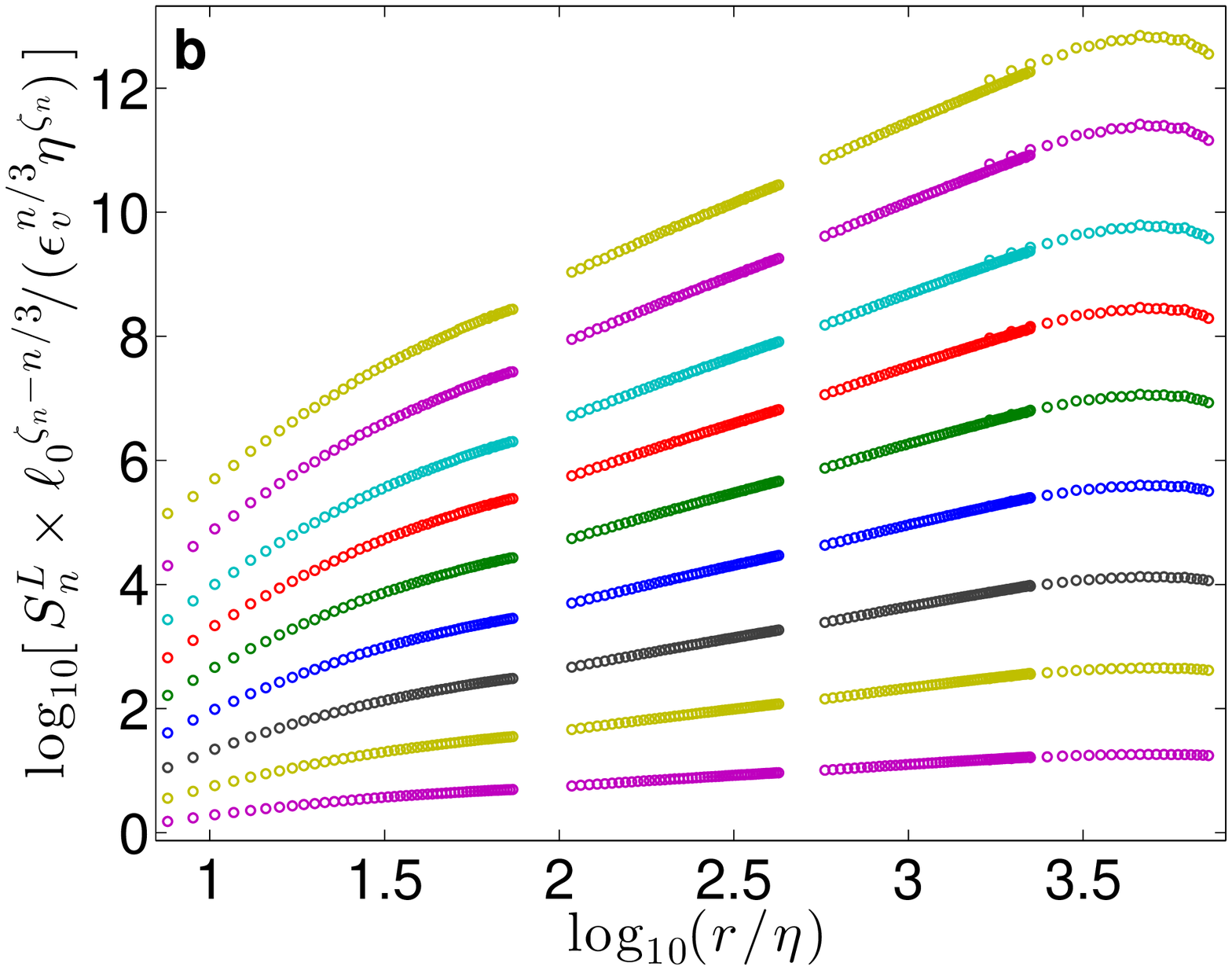}
\includegraphics[width=.495\textwidth]{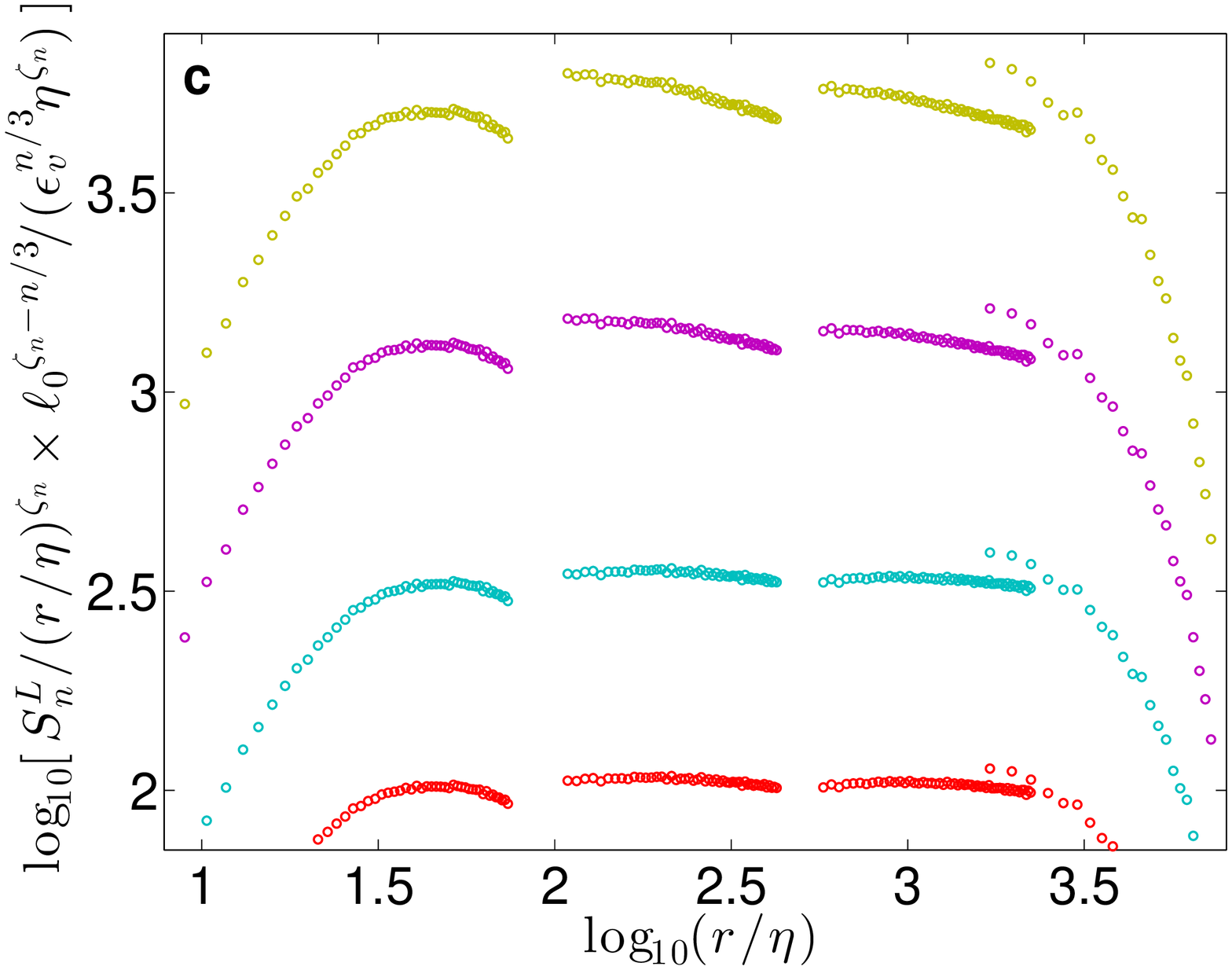}
\includegraphics[width=.495\textwidth]{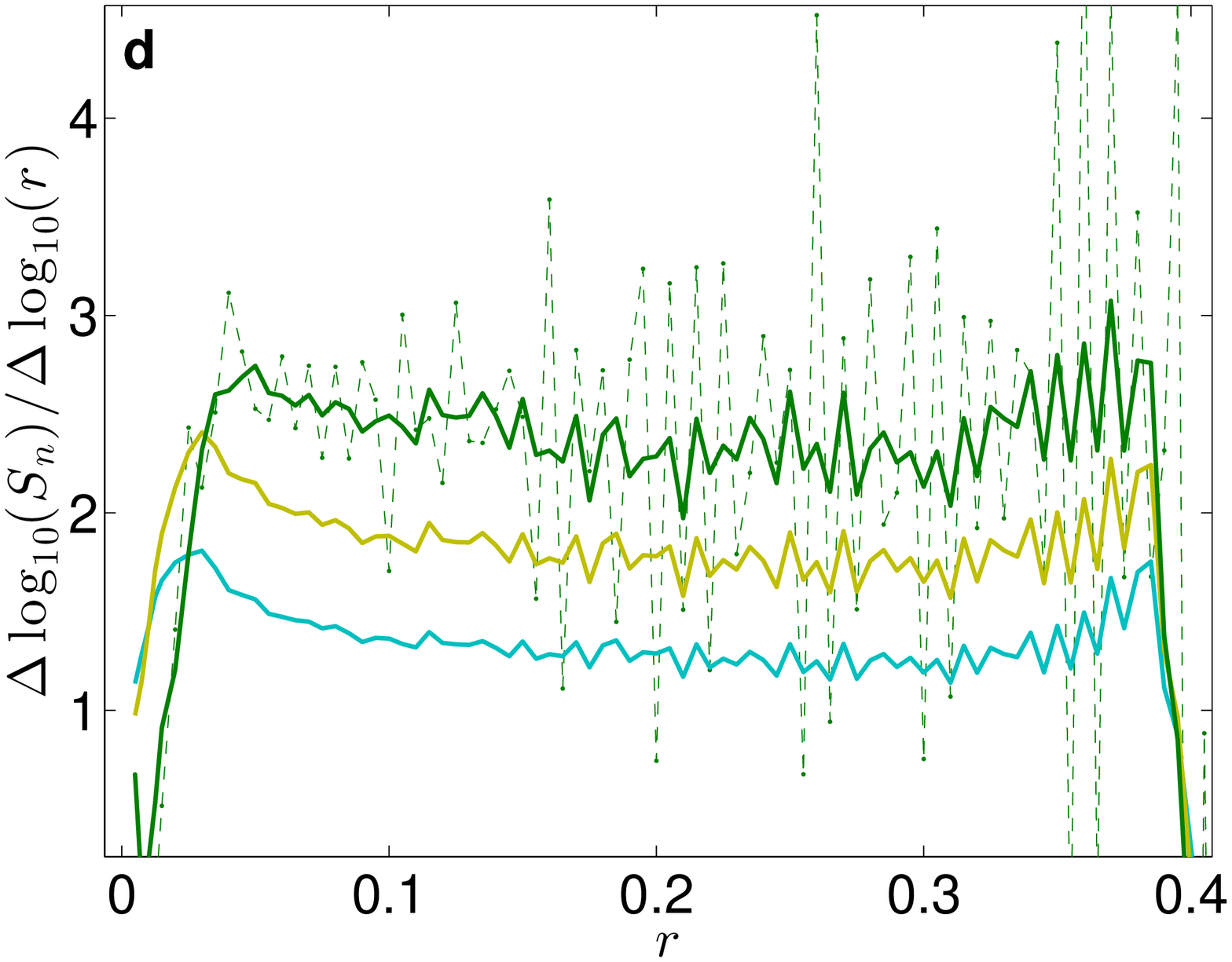}

\centering
\caption{\label{SpIntermNoCropAndCrop} Longitudinal structure functions (up to 9th order) rescaled using the intermittent scaling relation. a) plots showing all data points from all cases (A though D).  b)  Cropped structure functions where in each cases, the segment of the data affected by measurement noise (small $r$) and by finite measurement volume (large $r$) are removed. \ew{c) Same as b) but only showing $S^L_6$ to $S^L_9$ (from bottom to top) and each curve is compensated by the global $\zeta_n$ given in Table~\ref{table:zetap}. d) Local slopes, $\Delta log_{10}(S_n)/\Delta log_{10}(r)$  versus $r$ for case B. Solid lines from top to bottom are for $n= 9, 6, 4$ respectively and they are smoothed by running average method. The dash line is the respective un-smoothed data. The curve fitting range is selected as the intermediate part by removing the strongly varying or fluctuating parts at small and large $S_3$.}}
\vspace{20pt}
\end{figure}

Local velocity measurements are performed using SPIV, providing the radial, axial and azimuthal velocity components on a meridional plane of the flow through a time series of $30,000$ independent time samples. In the sequel, we work with dimensionless quantities, using $R$ as the unit of length, and $(2\pi F)^{-1}$ as the unit of time, $F$ being the rotational frequency of the impellers. Formally, since we use $50\%$ overlapping interrogation box, the spatial resolution of our measurement is twice the grid spacing $\delta x$, which depends on the cameras resolution, the field of view and the size of the windows used for velocity reconstruction. In the sequel, we use 2M-pixels cameras at two different optical magnifications, to get \ew{ one set of measurements with field of view covering the whole space between the impellers with area of roughly $20\times20$ cm$^2$ ($\delta x=3.4$ mm, $32\times 32$ pixels interrogation windows),  and three sets with field of view of $4\times4$ cm$^2$ centered at the center of the experiment (case A and C with $16\times 16$ pixels windows, $\delta x=0.24mm$,  and case B with $32\times 32$ pixels windows, $\delta x=0.48mm$). The velocities measured using PIV have uncertainties due to random fluctuations (fluctuating number of particles in each interrogation window, optical noise etc) and averaging error (velocities are smoothed over interrogation window, thickness of laser sheet). These may result in unreliable calculation of velocity differences at the smallest distances. Thus velocity difference at the smallest distances are removed from further analysis (more on this the sequel).}
%The PIV velocity reconstruction procedure induces uncertainty, the intensity of which decreases with the width of the interrogation windows. The influence of the uncertainty is most severe at locations where the velocity or its differences are small, thus our results at smallest spatial difference are not reliable and are removed (more on this the sequel).} 
Since we are interested only in statistics of the velocity field in the inertial subrange of turbulence, we remove the large scale inhomogeneous artifact of the swirling flow by subtracting the long time average from each instantaneous velocity field. \ew{Using the in-plane components of the velocities and spatial separations}, we then compute the velocity increments as $\delta \bm{ u} (\bm{ r}) =\bm{ u}(\bm{ x}_{} + \bm{ r}_{}) - \bm{ u}(\bm{ x}_{})$,  $\bm{ x}^{}$ and $\bm{ r}^{}$ being the position and spatial increment vectors in our measurement plane.
From this, we obtain the longitudinal structure functions via the longitudinal velocity increment $\delta u_L(r)=\delta \bm{ u} (\bm{ r})\cdot \bm{ r}^{}/{r}^{}$:
\begin{equation}
S_n(r)=\langle (\delta u_L)^n\rangle,
\label{strucfonc}
\end{equation}
where $\langle \rangle$ means average over time, \ew{all directions and the whole view area}. Our statistics therefore includes around $10^9$ to $10^{10}$ samples
(depending on the increment length), allowing convergence of structure functions up to $n=9$ in the inertial ranges (more details below). The structure functions ($S_n$) are shown in Fig.~\ref{SpIntermNoCropAndCrop}, where we have conjoined the four case (A through D) by rescaling the structure function in each case as $S_n\times {\ell_0}^{\zeta_n-\sfrac{n}{3}}/({\epsilon_v}^{\sfrac{n}{3}}\, \eta^{\zeta_n})$ and the abscissa as $r/\eta$ where $\zeta_n$ is the corresponding scaling exponents of the structure functions in the manner: $S_n(r)=C_n(\epsilon r)^{n/3}(r/\ell_0)^{\zeta_n -n/3}$ , with $\ell_0$ the characteristic large scale of the flow which we take as equals to the radius of the impellers (more on the computation of $\zeta_n$ and the choice of $\ell_0$ in Section \ref{exponents}). We note that in doing so, we have used the more general form of scaling relation for the structure functions that takes into account turbulent intermittency (as described earlier), with the K41 theory recovered if $\zeta_n=n/3$. As illustrated in Fig.~\ref{SpIntermNoCropAndCrop}a, in each segment of the $S_n$ curve represented by a single color, the behavior of $S_n$ at its large scale limit is altered by finite size effects, while at the small scale ends they are polluted by measurement uncertainties or lacks statistical convergence. \ew{We thus remove these limits and keep only the intermediate power-law-like segments for the analysis of scaling exponents in the sequel, the results are displayed in Fig.~\ref{SpIntermNoCropAndCrop}b and \ref{SpIntermNoCropAndCrop}c (note: for illustrative purpose here we also show the dissipative scales of case A and largest scales of case D which would correspond to large eddy scale). Specifically, data cropping is done by inspecting the, albeit noisy, local slopes plots i.e. $\Delta log_{10} (S_n) / \Delta log_{10}(r)$ versus $log_{10}(r)$ as exemplify by Figure~\ref{SpIntermNoCropAndCrop}d. We remove the strongly varying or fluctuating parts at small and large $r$ (based on smoothed data). However in case B and C, further removal of points at small $r$ were peformed in view of unsatisfactory statistical convergence (see Discussions for details on data convergence). The retained ranges are respectively for case A to D, $log_{10}(r/\eta) \in (1.5, 1.87), (2.02, 2.63), (2.75, 3.35), (3.2, 3.35).$}
%$log_{10}(r/\eta) \in (0.8, 1.87), (2.02, 2.63), (2.75, 3.35), (3.2, 4).$}
%We note that in these segments, the statistical convergence of the moments improves significantly with $r$ and deteriorates with order of the moments (see Discussions for details on data convergence).

%In the worst case (9th order, smallest $r$ in each segments), the uncertainty of the moments are about $1\%$ (this is estimated by first extrapolating the tails in the graph of $\delta u$ versus $\delta u^9P(\delta u)$ using exponential lines, $P(\delta u)$ being the probability density function; followed by evaluating the area-under-curve of the part of the exponential tails that are unresolved by our data, thus giving an estimate of the uncertainty). 
In the sequel, we discuss the computations of the kinetic energy dissipation rates and scaling exponents ($\zeta_n$) used in Fig.~\ref{SpIntermNoCropAndCrop}.

%One sees that at small orders, $n$, we obtain a smooth and continuous function, whilen
%jumps" in between two cases appear as $n$ grows larger and larger, showing that our normalization is not consistent in between different cases. As discussed in the introduction, this is due to two effects: i) the fact that the local dissipation rate $\epsilon_v$ differs from $\epsilon_g$ due to flows inhomogeneities; ii) intermittency corrections. The two effects are discussed and corrected in the 
%sequel.

%\begin{figure}
%\includegraphics[width=.49\textwidth]{figures/MomentsGlobalNoCrop.eps}
%\includegraphics[width=.49\textwidth]{figures/Moments_rescaled_Anti_Crop.eps}
%\centering
%\caption{\label{SpRawAndCrop} Left: Raw structure functions normalized by $\epsilon_g$. Right:  Cropped raw structure functions, normalized by $\epsilon_g$}
%\label{bkmabc}
%\vspace{20pt}
%\end{figure}

\subsection{Determination of local energy dissipation rates}

We use the local average kinetic energy dissipation rates, $\epsilon_v$, to rescale the structure functions. For this, we need accurate measurements of $\epsilon_v$. We determine $\epsilon_v$  for the four cases in two steps. In step one, we first determine $\epsilon_v$ in case A, where our data span both the dissipative and the lower inertial scales of turbulence, by constraining the value of $\epsilon_v$ such that both scaling laws of the second order structure function in the inertial subrange (K41) i.e. $S_2(r) = C_2 (\epsilon r)^{2/3}$ and the dissipative scale i.e. $S_2(r) = (\epsilon/15\nu)r^2$ are well satisfied. $C_2$ is the universal Kolmogorov constant with a nominally measured value of $C_2=2$ \cite[see e.g.][]{Pope00}. We note that the dissipative scaling formula implicitly assumes that the average dissipation rate can be replaced by its one-dimensional surrogates, which is expected to be accurate when at least local statistical isotropy is satisfied by the turbulent flow as in our case.

\begin{figure}
\includegraphics[width=.495\textwidth]{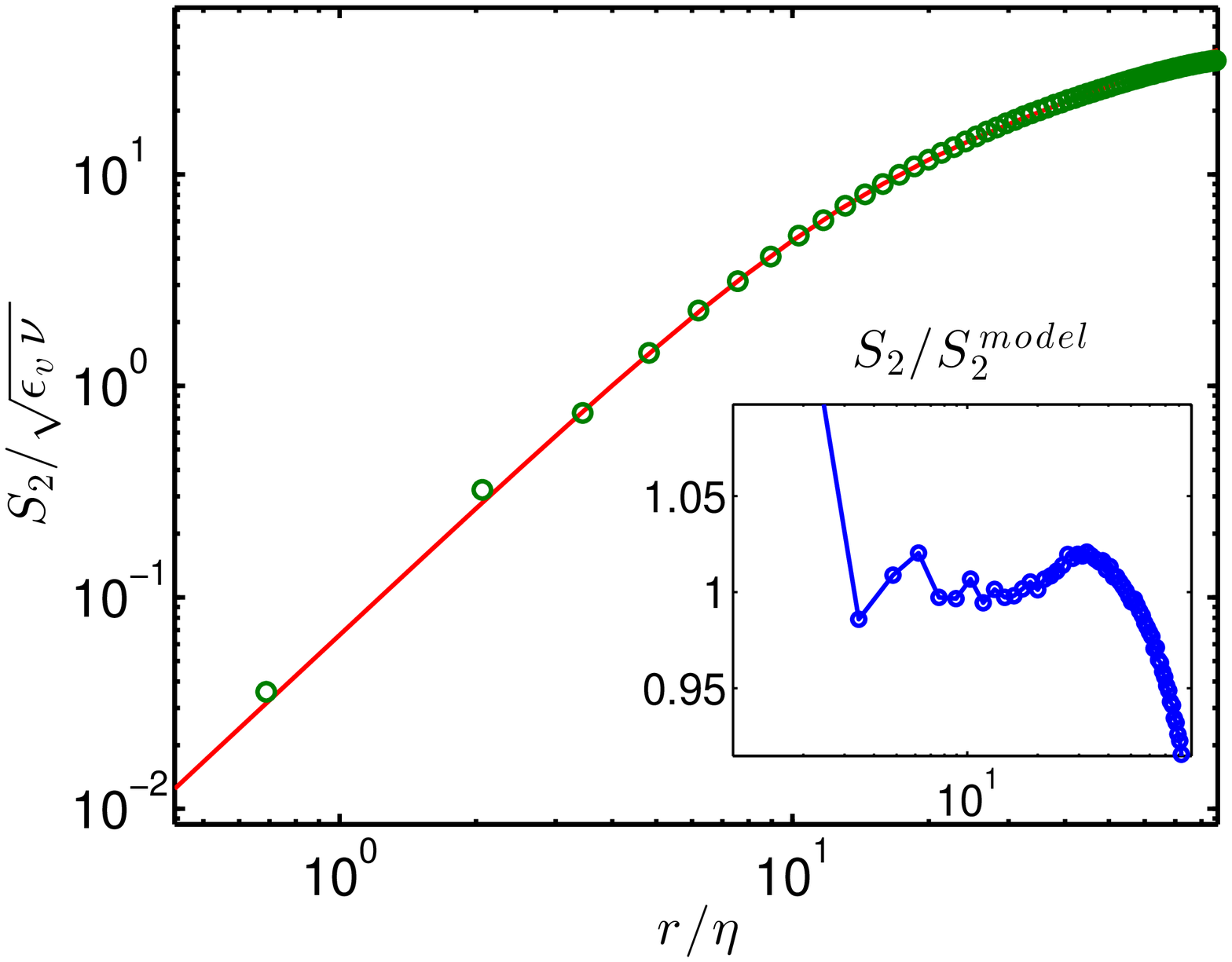}
\includegraphics[width=.495\textwidth]{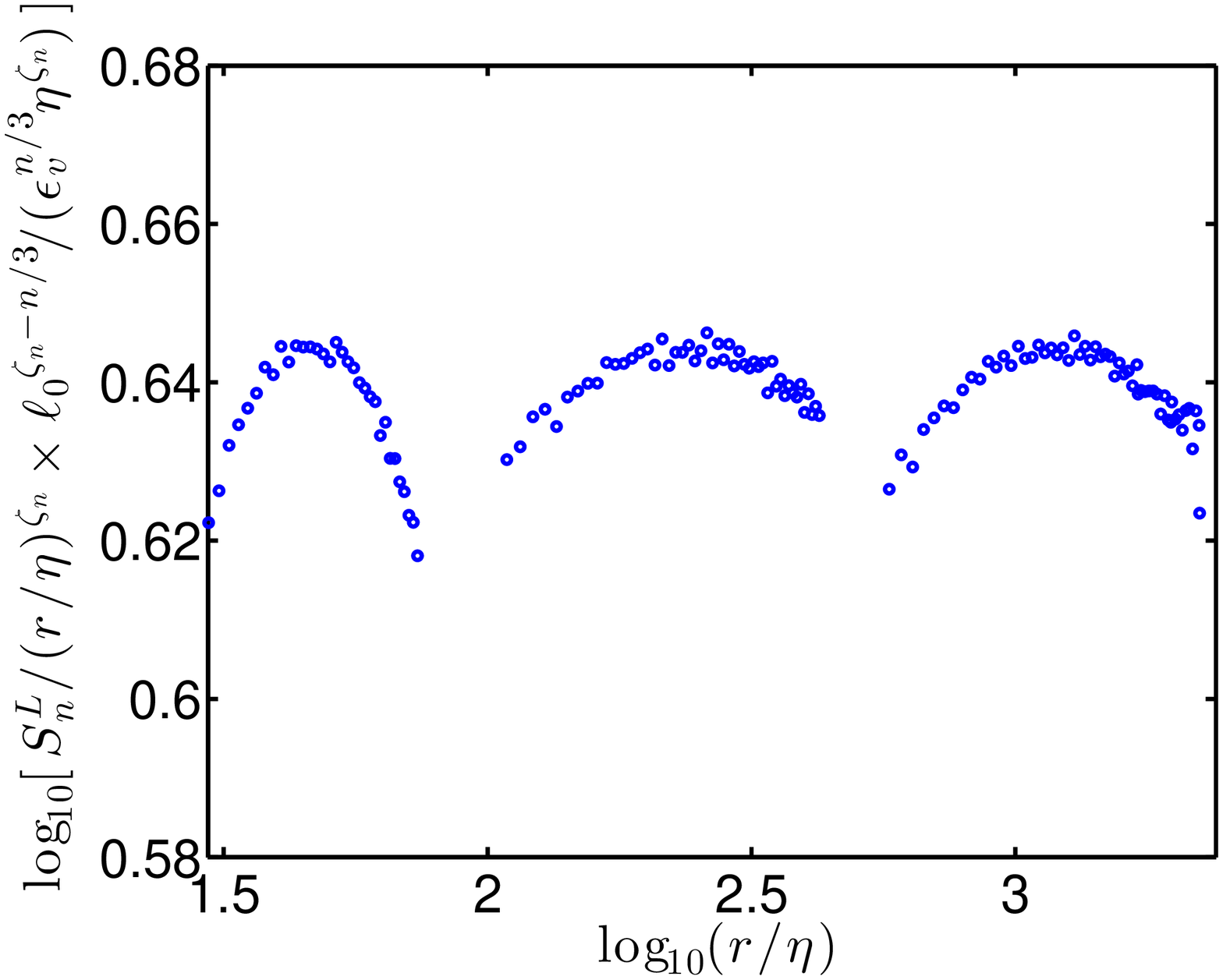}

\centering
\caption{\label{S2_dissip} Left: Nondimensionalized second order structure function ($S_2/(\epsilon_v \nu)^{1/2}$)  from case A. Comparison with the Sirovich model. Inset: discrepancy between the two curve (ratio of $S_2$ to the model). Close agreement between the two implies an accurate estimates of $\epsilon_v$ (for case A). The experimental noise in $S_2$  is partially removed by subtracting a value of 0.2. Right: Third order structure function compensated by power 1.}
\vspace{20pt} 
\end{figure}

A convenient way to achieve this is \ew{ by tuning value of $\epsilon_v$} in order to match our $S_2$ against the form $S_2=\alpha\,r^{2/3}/[1+(r/r_c)^{-2}]^{2/3}$ that contains the correct asymptotic both in the inertial and dissipative limits \ew{(this would be stronger than e.g. estimating $\epsilon_v$ using inertial sub-range data alone)}. This form was originally obtained by \cite{Sirovich1994} using Kolmogorov relation for third order structure function \citep{K41b} ( 4/5-law with exact viscous correction). The constants are further determined by asymptotically matching to the above scalings laws, giving $\alpha= \epsilon^{2/3}C_2$ and $r_c = (15C_2\nu/\epsilon^{1/3})^{3/4} \equiv (15C_2)^{3/4}\eta$. Figure \ref{S2_dissip} shows the non-dimensionalized second order structure function of case A, $S_2/(\epsilon_v \nu)^{1/2}$ as function of $(r/\eta)$ compared the Sirovich form (similarly non-dimensionalized, with $C_2=2$) for comparison. One can see a good agreement between the two curves in both dissipative and inertial range as well as at the transition regime, with discrepancies below $5\%$ excluding the far ends where data is affected by measurement uncertainties and view volume edges. This gives us confidence with regard to the estimation of $\epsilon_v$ for case A.

In step two, knowing the value of $\epsilon_v$ for case A, we determined $\epsilon_v$ of the other cases by constraining (assuming) that the conjoined 3rd order structure function of all cases, should globally scales \ew{(determined by curve fitting)} as a power law with exponent $\zeta_3=1$, as is predicted by K41 and supported by experiments  \cite[e.g. ][]{Anselmet1984} and numerical simulations \cite[e.g. ][]{Ishihara09}. As such, we have made the same assumption as in the extended self similarity method of \citep{Benzi93}. \ew{Figure \ref{S2_dissip}-right panel shows the conjoined $S_3$ rescaled using the resulting $\epsilon_v$.  }

%\ew{We close this section by noting that, alternatively we could estimates $\epsilon_v$ of all cases solely based on apply the K41 inertial range scaling as described above, these values (not shown) are 6 to10$\%$ lower than we report here.}  

%There is no fitting here. The constants are determined by the Kolmogorov constant = 2  and the dissipative asymptotic (definition of 1D epsilon). This support strongly our estimate of epsilon using S2, since the estimate of S3 will be significantly lower. S2 is normalize $epsilon^2/3$  (not zeta!).

\begin{table}
 % \begin{center}
\def~{\hphantom{0}}
 \begin{tabular}{lccccc}% 
    &$\zeta_1$&$\zeta_2$&$\zeta_4$ &$\zeta_5$&$\zeta_6$	\\
%\hline \hline
\rule{0pt}{3ex}
Arneodo ESS &$0.35\pm 0.03$&$0.7 \pm 0.03$ &$1.28\pm 0.03$ &$1.55 \pm 0.05$ &$ 1.77 \pm 0.05  $	 \\
\rule{0pt}{3ex}
%Belin et al. &$0.4 $&$0.7 $ &$ 1.26 $ &$1.50 $ &$ 1.71  $	 \\
%\rule{0pt}{3ex}

This paper, ESS & $ 0.36\substack{+0.005 \\ -0.005} $ & $ 0.69\substack{+0.005 \\ -0.005} $ &$1.29\substack{+0.005 \\ -0.005} $&$ 1.55\substack{+0.01 \\ -0.01} $ &$ 1.80\substack{+0.02 \\ -0.02} $	\\
\rule{0pt}{3ex}
This paper, global & $ 0.35\substack{+0.03 \\ -0.04} $ & $ 0.68\substack{+0.03 \\ -0.03} $ & $ 1.30\substack{+0.03 \\ -0.04} $ & $ 1.58\substack{+0.03 \\ -0.04} $ &$ 1.83\substack{+0.03 \\ -0.04} $	\\
 \rule{0pt}{8ex}

&$\zeta_7$&$\zeta_8$&$\zeta_9$\\
%\hline \hline
\rule{0pt}{3ex}
Arneodo ESS  &$2.03\pm 0.05 $ &$2.2 \pm 0.08$ &$2.38\pm 0.05$\\
\rule{0pt}{3ex}
%Belin et al. &$ 1.90 $&$ 2.08 $ &$ 2.19 $ 	 \\
%\rule{0pt}{3ex}

This paper, ESS &$ 2.02\substack{+0.03 \\ -0.03} $&$ 2.23\substack{+0.03 \\ -0.04} $&$ 2.41\substack{+0.05 \\ -0.05}$\\
\rule{0pt}{3ex}
This paper, global &$ 2.05\substack{+0.03 \\ -0.03} $&$ ^*2.35\substack{+0.04 \\ -0.05} $&$ ^*2.57\substack{+0.04 \\ -0.05}$\\
\rule{0pt}{3ex}

\end{tabular}
\caption{Comparison of measured scaling exponents between this paper and previous experiments. Arneodo ESS: results from various open turbulent flows in \cite{Arneodo96}.
% Belin et al.: swirling flow result in \cite{Belin1996}. 
This paper, ESS: results from this paper using ESS. This paper, global: results from this paper based on conjoined structures functions obtained at different Reynolds numbers. The values at highest orders marked with``$*$" are likely unreliable due to stark inconsistency with local scalings of the corresponding structure functions (details in text). The values of $\zeta_3$ in not shown since it is by assumption of the methodologies equals to unity in all cases.}
\label{table:zetap}
%\end{center}
\end{table}

\subsection{Determination of scaling exponents: ESS and global conjoin method}
\label{exponents}
\begin{figure}
\includegraphics[width=.45\textwidth]{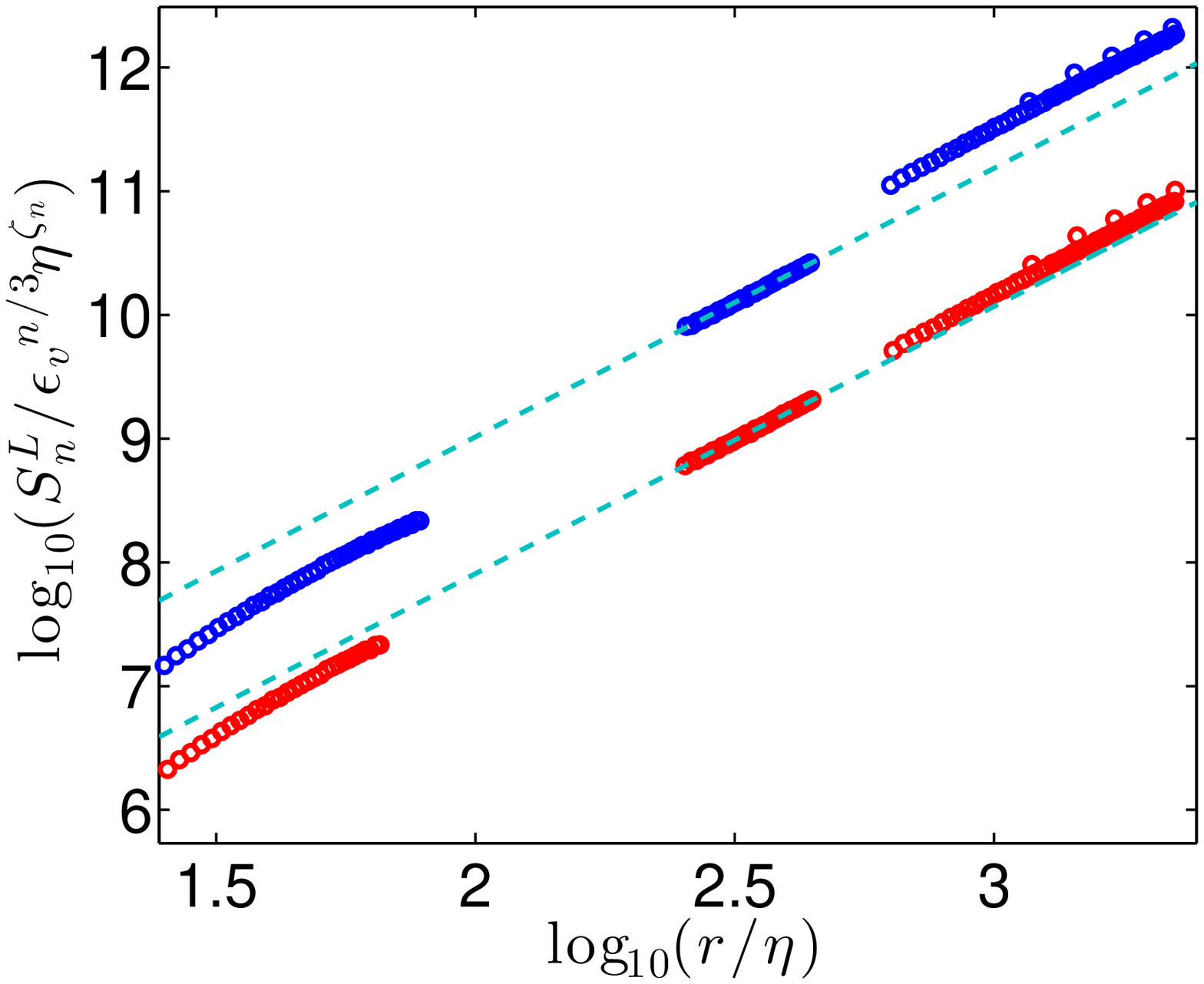}
\includegraphics[width=.45\textwidth]{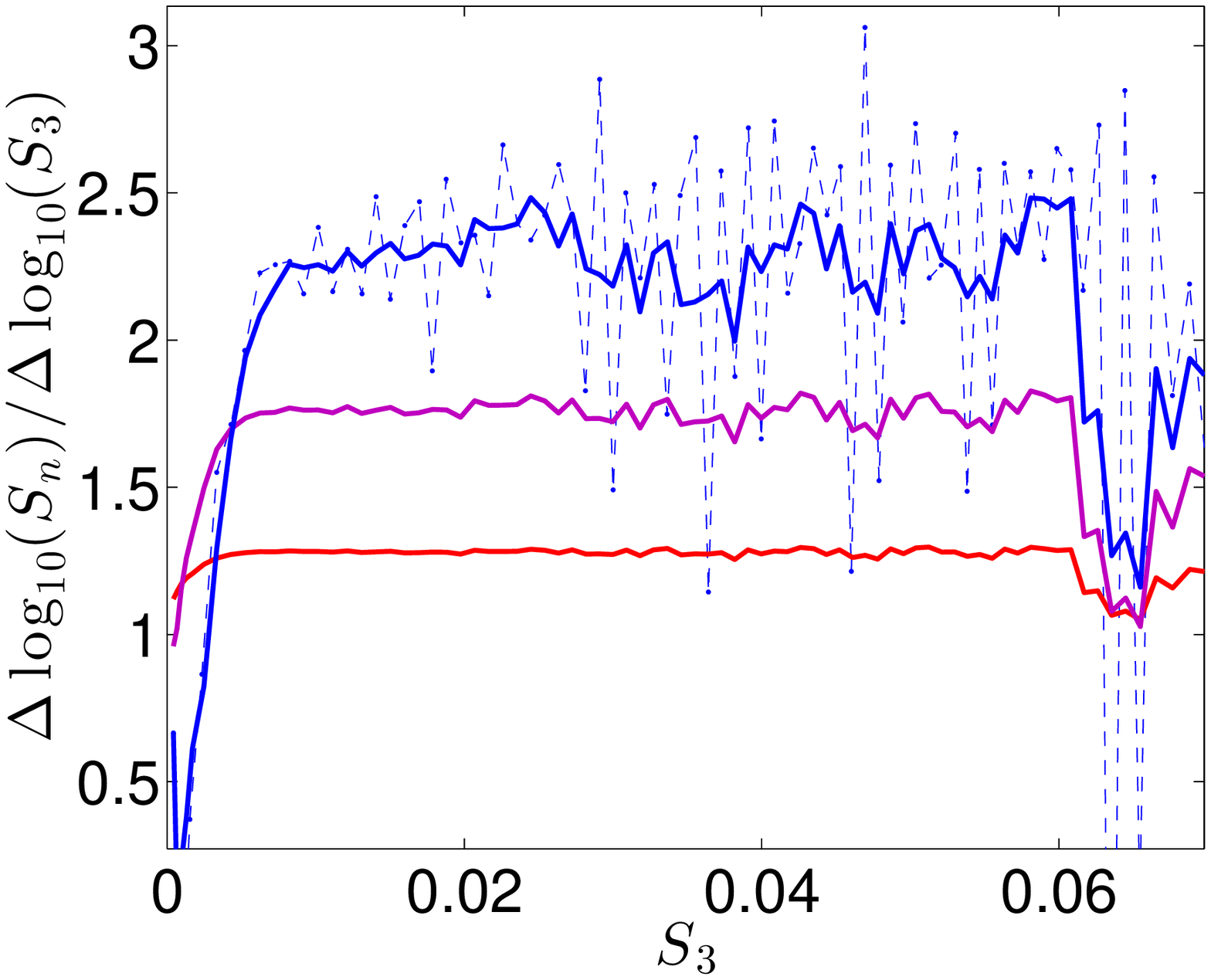}
\centering
\caption{\label{K41vsInterm} Left: Comparing consistency of structure functions rescaled using K41 and intermittent scaling relations. From top to bottom: blue circles is the 8th order structure function ($S_8$) rescaled using the K41 scaling ($\zeta_8=8/3 \approx 2.667$); red circles is $S_8$ rescaled via the intermittent relation with $\zeta_8=2.35$. Cyan dash lines are best fits to each set of data in the range $r/\eta=2.4-2.6$. The intermittent case is found to show higher general level of consistency (continuity) among the higher order structure functions (the difference is insignificant at lower orders). \ew{Right: Local slopes, $\Delta log_{10}(S_n)/\Delta log_{10}(S_3)$  versus $S_3$ for case B. Solid lines from top to bottom are for $n= 9, 6, 4$ respectively and they are smoothed by running average method. The dash line is the respective un-smoothed data. The ESS curve fitting range is selected as the intermediate part by removing the strongly varying or fluctuating parts at small and large $S_3$.}}
\vspace{20pt} 
\end{figure}

{
In Table \ref{table:zetap}, we report the scaling exponents ($\zeta_n$) of the structure functions by two different methods. Firstly, we apply the extended self similarity (ESS) method  \citep{Benzi93} to each of the four experimental cases (A through D). This essentially involves plotting $S_n(r)$ versus $S_3(r)$ in logarithmic axes followed by curve fitting. \ew{The uncertainty of each $\zeta_n$ is given as the $95\%$ confidence interval of the least-square fitting algorithm. The fitted ranges are chosen by inspection of the, albeit noisy, local slopes plots i.e. $\Delta log_{10} (S_n) / \Delta log_{10}(S_3)$ versus $log_{10}(S_3)$ as exemplify by Figure~\ref{K41vsInterm}-right. In general, the range is selected as the intermediate part by removing the strongly varying or fluctuating parts at small and large $S_3$. However in case B and C, further removal points small $S_3$ are prompted by unsatisfactory statistical convergence (see Discussions for details on data convergence). Numerically the ranges are, respectively for case A to D, $log_{10}(S_3) \in (-1.88, -1.5), (-2.0, -1.2), (-1.9, -1.1), (-1.67, -1.21).$}
%To estimate the uncertainties, the lower (upper) bound of each $\zeta_n$ measurement is obtained by a separate curve fit to the lower (upper) sub-range of the full fitting window, with each sub-range width of 0.2 decade in the abscissa. 
Since this produces four independent measurements of $\zeta_n$ for each order (owing to the 4 cases), we report their averages in Table \ref{table:zetap}.  
}

Secondly, we conjoin the non-dimensionalized structure functions (using  $\epsilon_v$ and $\eta$) from the four cases and apply curve-fitting to the combined structure functions to obtain the global estimates of $\zeta_n$. As shown in Fig. \ref{K41vsInterm}-left, we found that structure functions join significantly better when they are rescaled based on the scaling relation that takes into account intermittency, namely: $S_n\times {\ell_0}^{\zeta_n-\sfrac{n}{3}}/({\epsilon_v}^{\sfrac{n}{3}}\, \eta^{\zeta_n})$ versus $r/\eta$ (as discussed above).
%$S_n(\hat{r})=C_n \eta^{\zeta_n}(\epsilon \hat{r})^{n/3}(\hat{r}/\ell_0)^{\zeta_n -n/3}$, where $\hat{r}=r/\eta$ and $\ell_0$ is the characteristic large scale of the flow. 
We take $\ell_0=R$ (equals the impeller radius) in the current analysis, any global refinement in the magnitude of $\ell_0$ will not affect our estimate of $\zeta_n$, as it only multiply all $S_n$ by a constant factor. Specifically, the combined structure functions thus rescaled, exhibit significantly better continuity as compared to their K41 scaled counterparts. The non-dimensional structure functions $S_n(\hat{r})\times{\ell_0}^{\zeta_n-\sfrac{n}{3}}/({\epsilon_v}^{\sfrac{n}{3}}\, \eta^{\zeta_n})$, as such are dependent on values of $\zeta_n$, thus in order to improve accuracy, we iterate between rescaling and curve fitting to arrive at a set of self consistent $\zeta_n$. However, we observe that the self-consistency of this method gradually deteriorate at higher orders, essentially giving global iterated $\zeta_n$ values that are highly inconsistent with their piecewise estimates. One plausible cause of this could be the possibility that the relevant external scale $\ell_0$ varies between the different set of experiments. We, unfortunately, do not have way to independently measure $\ell_0$, but we note that allowing $\ell_0$ to vary up to 20$\%$ could remove such inconsistency. In view of this, for the 8th and 9th order, such self-iterative results are less reliable, hence our best estimates for $\zeta_8$ and $\zeta_9$ should still be the EES results. 

\subsection{Transversal scaling exponents}
\ew{While the main focus of the current paper is on the comparison of longitudinal structure functions, we briefly present the results on transversal structure functions here. Unlike their longitudinal counterparts, past results on scaling exponents of the transversal structure functions ($\zeta^T_n$) do not inspire strong consensus. There were conflicting results on whether they are equivalent to the longitudinal ones and some works suggested that they might depend significantly on large scale shear (for details see e.g. discussion by \cite{Iyer2017} and reference therein). However, the recent results of \cite{Iyer2017} strongly suggests that, when large scales inhomogeneities are absent, the two sets of exponents (longitudinal and transversal) are equivalent and subject to a single similarity hypothesis. Their findings also implies that previous conflicting results could be explained by the relatively much slower approach of the higher order transversal exponents to their ultimate large Reynolds number limits and possibly by presence of large scale shear. It remains for more works to substantiate this important finding. In view of this, here we present only the ESS results on the transversal exponents, as their strong dependence on Reynolds precludes any attempts to conjoin them using our global method described previously. The ESS results is shown in Table~\ref{table_zeta_t}. The values for $\zeta^T_n$  at higher orders are lower than their longitudinal counterparts, consistent with some previous experiments e.g. \citep{Dhruva97,Shen2002}. Our results shows weak evidence that $\zeta^T_n$ increases with Reynolds number for $n>6$. }

\begin{table}
  \begin{center}
\def~{\hphantom{0}}
 \begin{tabular}{cccccc}%
 %\vspace{2pt}
 Case & A & B & C & D & Average	\\
 \hline
 \vspace{3pt}
 $Re$ & $6\times 10^3$  & $6\times 10^4$ & $3\times 10^5$ & $3\times 10^5$	\\
%\hline 
\vspace{3pt}
$\zeta^T_1$ &$ 0.40 $&$ 0.38  $&$ 0.39 $&$ 0.39 $ &$ 0.39 $\\ 
\vspace{3pt}
%\hline
$\zeta^T_2$ &$ 0.73 $&$ 0.71 $&$ 0.72 $&$ 0.72 $&$ 0.72 $\\
\vspace{3pt}
%\hline
$\zeta^T_4$ &$ 1.22 $&$ 1.25 $&$ 1.24 $&$  1.23-1.24  $&$ 1.24 $\\
\vspace{3pt}
%\hline
$\zeta^T_5$ &$ 1.40-1.41 $&$ 1.45 $&$ 1.43-1.44 $&$ 1.43  $&$ 1.43 $ \\
\vspace{3pt}
%\hline
$\zeta^T_6$ &$ 1.54-155 $&$ 1.62 $&$ 1.59-1.60 $&$  1.58-1.60  $&$ 1.59 $\\
\vspace{3pt}
%\hline
$\zeta^T_7$ &$ 1.64-1.67 $&$ 1.76 $&$ 1.73-1.74 $&$ 1.69-1.73 $&$ 1.72 $\\
\vspace{3pt}
%\hline
$\zeta^T_8$ &$ 1.72-1.75 $&$ 1.85-1.86 $&$ 1.83-1.86 $&$ 1.77-1.84 $&$ 1.81 $\\
\vspace{3pt}
%\hline
$\zeta^T_9$ &$ 1.78-1.81 $&$ 1.90-1.92 $&$ 1.92-1.95 $&$ 1.80-1.93 $&$ 1.88 $\\

%\hline

\end{tabular}

\caption{ Scaling exponents of the transversal structure functions of velocity. $\zeta^T_n$ values that are not presented as a ranges imply uncertainties of $\pm 0.005$, except the last column (Average) which are simple averages of the mid-values of each row. }
 \label{table_zeta_t}
\end{center}
\end{table}

\section{Discussions}

\begin{figure}
\centering
\includegraphics[width=.45\textwidth]{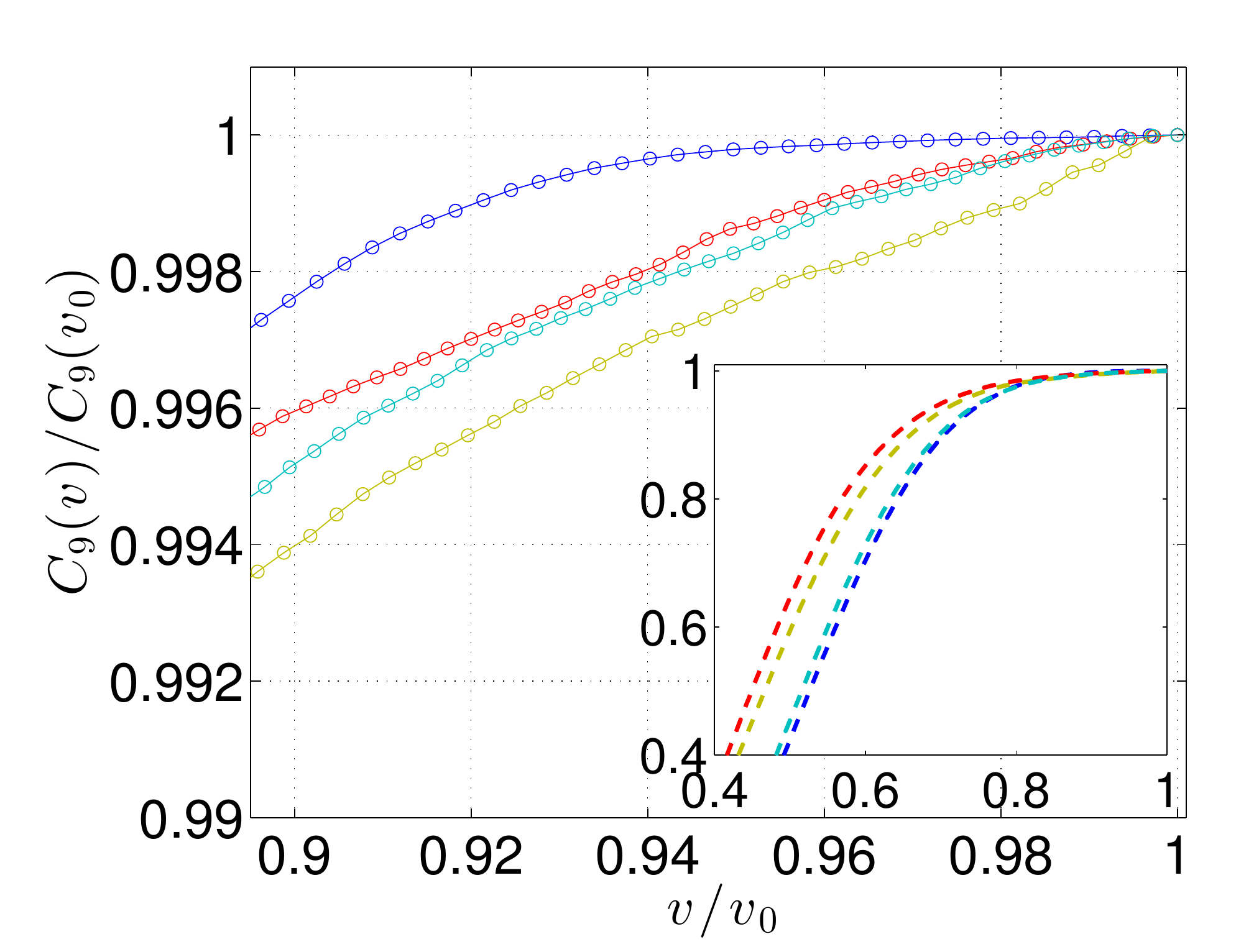}
\includegraphics[width=.445\textwidth]{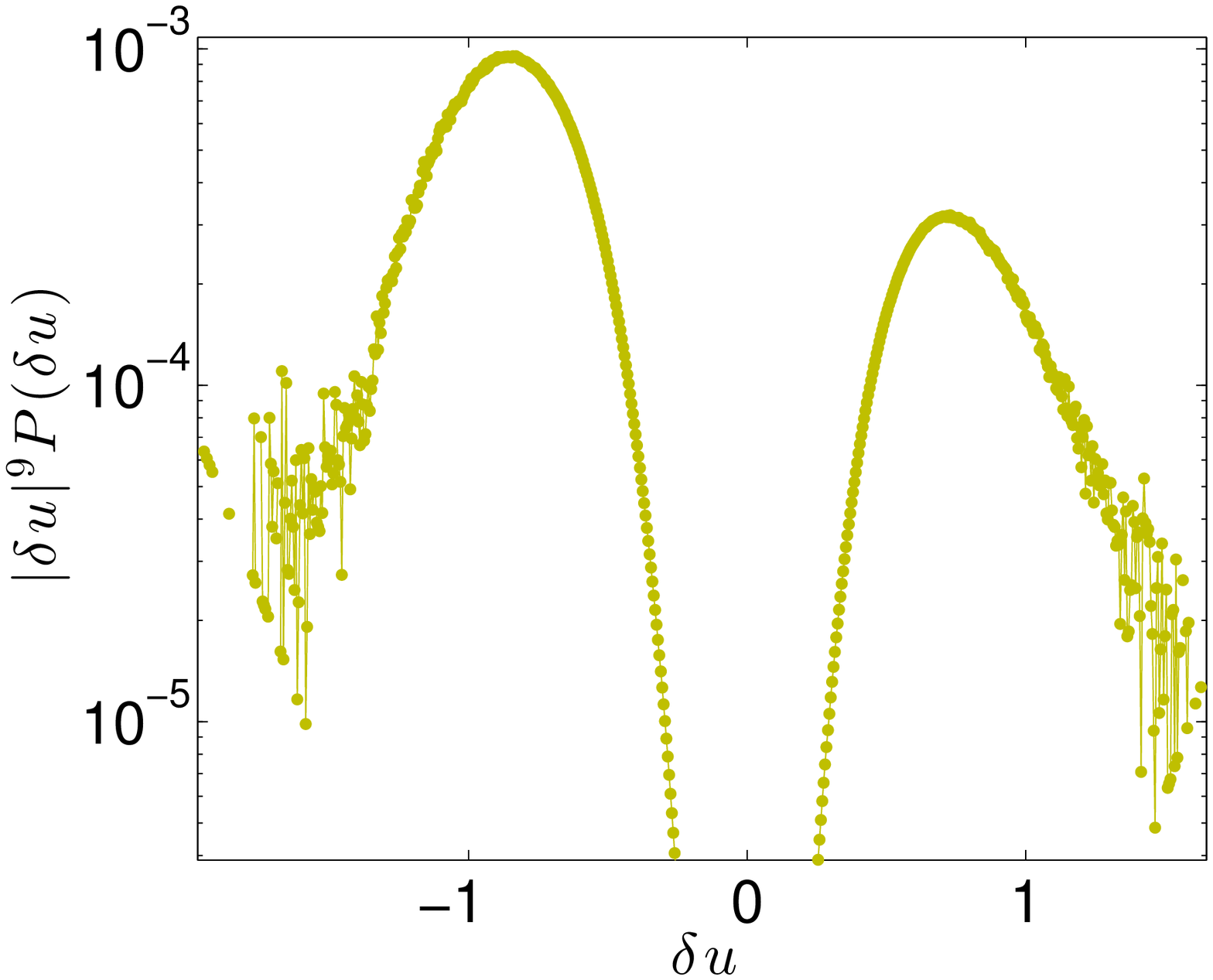}
\centering
\caption{\label{Converge} Statistical convergence of the structure functions. Left: main figure and the inset, we plot $C_9(v)/C_9(v_0)$ versus $v/v_0$, where $C_n(v)\int^{v}_0 v'^nP(v')dz'$, $v$ and $v'$ are modulus of velocity differences, $P$ the PDF and $v_0$ is the value of $v$ where $P(v)$ first reach zero value in our data. The curves (blue, gold, red, cyan) represent points from case A to D respectively that are the least converged and that are used to calculate $\zeta_n$ (and retained in Fig.~\ref{SpIntermNoCropAndCrop}-right).  As the inset shows, $C_9(v)$ increase with $v$ before they start to saturate at large $v$. The main figure verifies that all data points used in our analysis of the scaling exponents will not vary more that $1\%$ even if additional sampling would increase $v_0$ by $10\%$. Right: $|\delta u|^9P(\delta u)$ versus $\delta u$ corresponding to case B (gold) in the left panel, where $\delta u$ is the velocity difference.}

\vspace{0pt} 
\end{figure}

\ew{\textit{Statistical convergence of data.} There exist various ways of characterizing statistical convergence of structure functions. We follow the method used in \citep{Gotoh02}, as it reveals directly the possible rate of change of the moments with respect to increasing statistics. This involves plotting $C_n(v)\int^{v}_0 v'^nP(v')dz'$ where $v$ and $v'$ are modulus of velocity differences and $P$ the corresponding PDF. Figure~\ref{Converge} represents $C_9(v)/C_9(v_0)$ as function of $v/v_0$, where $v_0$ is the 1st value of $v$ at which $P(v)=0$ in our data. The four curves represent the least converged points in each case (A to D) used in our calculation of $\zeta_n$ and are respectively at $log_{10}(r/\eta)= 1.51, 2.04, 2.76, 3.23$. They are also
points of smallest $r$ retained in each cases, since convergence improves with $r$. To ease comparison with other works, we define our criterion for convergence as: $C_n(v)$ should not vary more than $1\%$ when $v_0$ is extended by $10\%$, consistent with what is shown in Figure~\ref{Converge}. }

\ew{ \textit{Flow inhomogeneity.} In this work, we attempt to minimize the influence of large scale inhomogeneity and anisotropy by subtracting the mean flow pattern from our data and average over all directions in the region near the symmetric center of the flow where the flow is roughly homogeneous. This however could not guarantee that all influence of inhomogeneity and isotropy has been removed, especially for case D where the view area is large. A full analysis of this issue will be the subject of future work.}

\section{Conclusion}
We use SPIV measurements of velocities in a turbulent von K\'arm\'an flow to compute longitudinal structure function up to order nine without using Taylor-hypothesis. Our multi-scale imaging provides the possibility to access scales of the order (or even smaller than) the dissipative scale, in a fully turbulent flow.  Using magnifying lenses and mixtures of different composition, we  adjust our resolution, to achieve velocity increment measurements spanning a range of scale between one Kolmogorov scale, to almost $10^{3.5}$ Kolmogorov scale, with clear inertial subrange spanning about 1.5 decades. Thanks to our large range of scale, we can compute the global scaling exponents by  analyzing conjoined data of different resolutions to complement the analysis of extended self-similarity. Our results on the scaling exponents ($\zeta_n$), where reliable, are found to match the values observed in turbulent flows experiments with open geometries  \citep{Anselmet1984, Stolovitzky1993, Arneodo96}, numerical simulations \citep{Gotoh02,Ishihara09} and the theory of \cite{She94}, in contrast with previous measurement in von K\'arm\'an swirling flow using Taylor-hypothesis, that reported  scaling exponents that are significantly smaller \citep{Maurer94, Belin1996}, which raised the possibility that the universality of the scaling exponents are broken with respect to change of large scale geometry of the flow. Our new measurements, that do not rely on Taylor-hypothesis, suggest that the previously observed discrepancy could be due to a pitfall in the application of Taylor-hypothesis on closed, non-rectilinear geometry and that the scaling exponents might be in fact universal, regardless of the large scale flow geometry. 
%\ew{Such result  would strengthen and generalize the conclusions of theoretical analysis and numerical simulation of Navier-Stokes equation that predict that the scaling properties of the structure functions are universal in the isotropic sector \cite{Biferale05}.}

%\begin{figure}
% % \centerline{\includegraphics{trapped}}% Images in 100% size
%  \caption{Trapped-mode wavenumbers, $kd$, plotted against $a/d$ for
%    three ellipses:\protect\\
%    ---$\!$---,
%    $b/a=1$; $\cdots$\,$\cdots$, $b/a=1.5$.}
%\label{fig:ka}
%\end{figure}
%
%\begin{figure}
%%  \centerline{\includegraphics{modes}}
%  \caption{The features of the four possible modes corresponding to
%  (\textit{a}) periodic\protect\\ and (\textit{b}) half-periodic solutions.}
%\label{fig:kd}
%\end{figure}
%
%\subsection{Tables}
%Tables, however small, must be numbered sequentially in the order in which they are mentioned in the text. The word \textit {table} is only capitalized at the start of a sentence. See table \ref{tab:kd} for an example.
%
%\begin{table}
%  \begin{center}
%\def~{\hphantom{0}}
%  \begin{tabular}{lccc}
%      $a/d$  & $M=4$   &   $M=8$ & Callan \etal \\[3pt]
%       0.1   & 1.56905 & ~~1.56~ & 1.56904\\
%       0.3   & 1.50484 & ~~1.504 & 1.50484\\
%       0.55  & 1.39128 & ~~1.391 & 1.39131\\
%       0.7   & 1.32281 & ~10.322 & 1.32288\\
%       0.913 & 1.34479 & 100.351 & 1.35185\\
%  \end{tabular}
%  \caption{Values of $kd$ at which trapped modes occur when $\rho(\theta)=a$}
%  \label{tab:kd}
%  \end{center}
%\end{table}

\par {\bf Acknowledgement}
This work has been supported by EuHIT, a project funded by the European Community Framework Programme 7, grant agreement no. 312778.

\bibliographystyle{jfm}
%\bibliography{jfm-instructions}

\begin{thebibliography}{}

\bibitem[Anselmet et al.(1984)]{Anselmet1984} Anselmet, F., Gagne, Y., Hopfinger, E. J., \& Antonia, R. A. (1984). High-order velocity structure functions in turbulent shear flows. Journal of Fluid Mechanics, 140, 63-89.

\bibitem[Arneodo et al.(1996)]{Arneodo96} ArnŽodo, A. E., Baudet, C., Belin, F., Benzi, R., Castaing, B., Chabaud, B., ... \& Dubrulle, B. (1996). Structure functions in turbulence, in various flow configurations, at Reynolds number between 30 and 5000, using extended self-similarity. EPL (Europhysics Letters), 34(6), 411.	

\bibitem[Belin et al.(1996)]{Belin1996} Belin, F., Tabeling, P., \& Willaime, H. (1996). Exponents of the structure functions in a low temperature helium experiment. Physica D: Nonlinear Phenomena, 93(1), 52-63.

\bibitem[Benzi et al.(1993)]{Benzi93}Benzi, R., Ciliberto, S., Tripiccione, R., Baudet, C., Massaioli, F., \& Succi, S. (1993). Extended self-similarity in turbulent flows. Physical Review E, 48(1), R29.

\bibitem[Biferale \& Procaccia (2005)]{Biferale05}Biferale, L., \& Procaccia, I.(2005). Anisotropy in turbulent flows and in turbulent transport. Physics Reports, 414.2,  43-164. 

\bibitem[Dhruva97 et al.(1993)]{Dhruva97}Dhruva, B., Tsuji, Y., \& Sreenivasan, K. R. (1997). Transverse structure functions in high-Reynolds-number turbulence. Physical Review E, 56(5), R4928.	

\bibitem[Gotoh et al.(2002)]{Gotoh02} Gotoh, T., Fukayama, D., \& Nakano, T. (2002). Velocity field statistics in homogeneous steady turbulence obtained using a high-resolution direct numerical simulation. Physics of Fluids (1994-present), 14(3), 1065-1081.

\bibitem[Huisman et al.(2013)]{Huisman2013} Huisman, S. G., Lohse, D., Sun, C. (2013). "Statistics of turbulent fluctuations in counter-rotating Taylor-Couette flows." Physical Review E 88.6: 063001.

\bibitem[Ishihara et al.(2000)]{Ishihara09}Ishihara, T., Gotoh, T., \& Kaneda, Y. (2009). Study of high-Reynolds number isotropic turbulence by direct numerical simulation. Annual Review of Fluid Mechanics, 41, 165-180.

\bibitem[Iyer et al.(2017)]{Iyer2017}Iyer, K. P., Sreenivasan, K. R., \& Yeung, P. K. (2017). Reynolds number scaling of velocity increments in isotropic turbulence. Physical Review E, 95(2), 021101.

\bibitem[Kolmogorov(1941)]{K41a} Kolmogorov, A. N. (1941a) "The local structure of turbulence in incompressible viscous fluid for very large Reynolds numbers." Dokl. Akad. Nauk SSSR. Vol. 30. No. 4.

\bibitem[Kolmogorov(1941)]{K41b} Kolmogorov, A. N. (1941b). Dissipation of energy in locally isotropic turbulence. In Dokl. Akad. Nauk SSSR (Vol. 32, No. 1, pp. 16-18).

\bibitem[Kolmogorov(1962)]{K62} Kolmogorov, A. N. (1962). A refinement of previous hypotheses concerning the local structure of turbulence in a viscous incompressible fluid at high Reynolds number. Journal of Fluid Mechanics, 13(01), 82-85.

\bibitem[Kuzzay(2015)]{Kuzzay15} Kuzzay, D., Faranda, D.  \& Dubrulle, B. (2015). Global vs local energy dissipation: The energy cycle of the von Karman flow.
 Phys. of Fluids,  27, 075105.

\bibitem[Lewis \& Swinney(1999)]{Lewis99} Lewis, G. S., and Swinney, H.L. (1999) "Velocity structure functions, scaling, and transitions in high-Reynolds-number Couette-Taylor flow." Physical Review E 59.5 : 5457.

\bibitem[Maurer et al.(1994)]{Maurer94} Maurer, J., Tabeling, P., \& Zocchi, G. (1994). Statistics of turbulence between two counterrotating disks in low-temperature Helium gas. EPL (Europhysics Letters), 26(1), 31.

\bibitem[Pinton \& Labb\'e (1994)]{Pinton94}  Pinton, J.-F. \& Labb\'e R. (1994). Correction to the Taylor hypothesis in swirling flows. J. Phys. II France, 4, 1461-1468. 

\bibitem[Pope(2000)]{Pope00} Pope, S. B. (2000). Turbulent Flows. Cambridge University Press.

\bibitem[Saw et al.(2016)]{Saw2016} Saw, E. W., Kuzzay, D., Faranda, D., Guittonneau, A., Daviaud, F., Wiertel-Gasquet, C., ... \& Dubrulle, B. (2016). Experimental characterization of extreme events of inertial dissipation in a turbulent swirling flow. Nature Communications, 7, 12466.

\bibitem[She \& Leveque(1994)]{She94}She, Z. S., \& Leveque, E. (1994). Universal scaling laws in fully developed turbulence. Physical Review Letters, 72(3), 336.

\bibitem[Shen \& Warhaft(2002)]{Shen2002}Shen, X., \& Warhaft, Z. (2002). Longitudinal and transverse structure functions in sheared and unsheared wind-tunnel turbulence. Physics of Fluids, 14(1), 370-381.

\bibitem[Sirovich et al.(1994)]{Sirovich1994} Sirovich, L., Smith, L., \& Yakhot, V. (1994). Energy spectrum of homogeneous and isotropic turbulence in far dissipation range. Physical Review Letters, 72(3), 344.

\bibitem[Sreenivasan \& Antonia(1997)]{Sreenivasan97} Sreenivasan, Katepalli R., \&  Antonia, R. A. (1997). The phenomenology of small-scale turbulence. Annual review of Fluid Mechanics 29.1: 435-472.

\bibitem[Stolovitzky et al.(1993)]{Stolovitzky1993} Stolovitzky, G., Sreenivasan, K. R., \& Juneja, A. (1993). Scaling functions and scaling exponents in turbulence. Physical Review E, 48(5), R3217.

\bibitem[Zocchi et al.(1994)]{Zocchi} Zocchi, G., Tabeling, P., Maurer, J., \& Willaime, H. (1994). Measurement of the scaling of the dissipation at high Reynolds numbers. Physical Review E, 50(5), 3693.

\end{thebibliography}

\end{document}